\documentclass[aps,prd,showpacs,twocolumn,
superscriptaddress]{revtex4-1}

\usepackage{graphicx}
\usepackage{dcolumn}

\usepackage{amsmath}
\usepackage{amssymb}
\usepackage{bm}
\usepackage{color}
\usepackage[normalem]{ulem}
\usepackage[dvipsnames]{xcolor}
\usepackage{hyperref}
\hypersetup{
colorlinks=true, 
linkcolor=blue,  
citecolor=cyan,  
}
\usepackage{multibib}

\newcommand{\bea}{\begin{eqnarray}}
\newcommand{\eea}{\end{eqnarray}}
\newcommand{\be}{\begin{equation}}
\newcommand{\ee}{\end{equation}}

\begin{document}
\title{
Observational appearance of Kaluza-Klein black holes
}

\author{Temurbek Mirzaev}
\email{mtemurbek22@m.fudan.edu.cn}
\affiliation{Center for Field Theory and Particle Physics and Department of Physics, Fudan University, 200438 Shanghai, China }

\author{Askar B. Abdikamalov}
\email{aaskar17@fudan.edu.cn}
\affiliation{School of Mathematics and Natural Sciences, New Uzbekistan University, Tashkent 100007, Uzbekistan}
\affiliation{Center for Field Theory and Particle Physics and Department of Physics, Fudan University, 200438 Shanghai, China }
\affiliation{Ulugh Beg Astronomical Institute, Astronomy Str. 33, Tashkent 100052, Uzbekistan }

\author{Ahmadjon A. Abdujabbarov}
\email{ahmadjon@astrin.uz}
\affiliation{Shanghai Astronomical Observatory, 80 Nandan Road, Shanghai 200030, China}
\affiliation{Ulugh Beg Astronomical Institute, Astronomy Str. 33, Tashkent 100052, Uzbekistan }
\affiliation{National University of Uzbekistan, Tashkent 100174, Uzbekistan}
\affiliation{Institute of Nuclear Physics, Tashkent 100214, Uzbekistan}
\affiliation{Institute of Fundamental and Applied Research, National Research University TIIAME, Kori Niyoziy 39, Tashkent 100000, Uzbekistan}

\author{Dimitry Ayzenberg}
\email{dimitry.ayzenberg@tat.uni-tuebingen.de}
\affiliation{Theoretical Astrophysics, Eberhard-Karls Universitat Tubingen, D-72076 Tubingen, Germany}

\author{Bobomurat Ahmedov}
\email{ahmedov@astrin.uz}
\affiliation{Ulugh Beg Astronomical Institute, Astronomy Str. 33, Tashkent 100052, Uzbekistan }
\affiliation{Institute of Fundamental and Applied Research, National Research University TIIAME, Kori Niyoziy 39, Tashkent 100000, Uzbekistan}
\affiliation{National University of Uzbekistan, Tashkent 100174, Uzbekistan}

\author{Cosimo Bambi}
\email[]{bambi@fudan.edu.cn}
\affiliation{Center for Field Theory and Particle Physics and Department of Physics, Fudan University, 200438 Shanghai, China }

\date{\today}

\begin{abstract}
The optical properties of rotating black holes in Kaluza-Klein theory described by the total mass, spin, and electric and magnetic charges are investigated in detail. Using a developed general relativistic ray-tracing code to calculate the motion of photons, shadows of Kaluza-Klein black holes are generated. The properties of the shadow and the light deflection angle around these black holes are also studied  in order to put constraints on the parameters of Kaluza-Klein black holes using M87* shadow observations. The possibility of imposing constraints on Kaluza-Klein black holes using shadow observations is investigated. Moreover, we find that small charges (electric and magnetic) of the black hole can meet these constraints. We conclude that with the current precision of the M87* black hole shadow image observation by the EHT collaboration, the shadow observations of Kaluza-Klein black holes are indistinguishable from  that  of  the  Kerr  black hole.  Much  better  observational accuracy than the current capabilities of the EHT  collaboration  are  required  in  order  to  place verified constraints on the parameters of modified theories of gravity in the strong field regime.
\end{abstract}

\maketitle

\section{Introduction}

The existing and rapidly growing experimental and observational data on astrophysical black hole candidates do not yet substantially support the idea that they should be well described by the Kerr spacetime metric. There are still open windows for consideration of other solutions within general relativity and/or modified theories of gravity~\cite{Bambi17c,Berti15,Cardoso16a,Cardoso17,Krawczynski12, Krawczynski18, Yagi16, Bambi16b, Yunes13, Tripathi2020yts}. Thus, one needs to think of how to test gravity theories through different observational and experimental data in both the weak and strong field regimes. 

Existing observational and experimental tests of gravity theories (see, e.g. \cite{Gravity18a}) cannot determine the metric coefficients of the Kerr spacetime with high accuracy. There is still hope that quickly developing technologies will lead to improved accuracy and help determine the physical characteristics of Kerr black holes with high confidence. The first ever image or so-called shadow of the supermassive black hole (SMBH) at the center of the galaxy M87 has been captured using the VLBI (Very Long Baseline Interferometry) technique by the Event Horizon Telescope (EHT) collaboration~\cite{EHT19a,EHT19b}. The obtained shadow of the SMBH M87* is consistent with expectations for the Kerr black hole shadow as predicted by general relativity. 
However, there is still a degeneracy problem since other axially-symmetric solutions within general relativity or modified theories of gravity can, in principle, mimic the Kerr spacetime parameters~\cite{Carballo-Rubio18,Abdikamalov19}.
Several papers have also shown that weakly naked singularities also mimic black holes very well~\cite{PhysRevD.65.103004, PhysRevD.77.124014}.

Particularly the shadows of axially-symmetric Kerr-like black hole solutions have been investigated and the deviation from the Kerr black hole has been analyzed~\cite{Takahashi05, Bambi09, Hioki09, Amarilla10, Bambi10, Amarilla12, Amarilla13, Abdujabbarov13c, Atamurotov13, Wei13, Atamurotov13b, Bambi15, Ghasemi-Nodehi15, Cunha15, Boshkayev15, Quevedo11, Javed19, Ovgun19,Ovgun19a, Johannsen11, Younsi16, Ayzenberg18}. Detailed studies of the observable quantities of the black hole shadow in various theories of gravity can be found in~\cite{Abdujabbarov15, Atamurotov15a, Ohgami15, Grenzebach15, Mureika17, Abdujabbarov17b, Abdujabbarov16a, Abdujabbarov16b, Mizuno18, Shaikh18b, Kogan17, Perlick17,Schee15, Schee09a, Stuchlik14, Schee09, Stuchlik10,Mishra19,Eiroa18, Giddings19, Ayzenberg18, Perlick_2022, PhysRevD.97.104062,PhysRevD.105.064056, Frost_2021, Manojlovic2020ndn, Olmo_2023, Vincent:2022fwj}.


Among the numerous alternative theories of gravity, the one proposed by Kaluza and Klein at the beginning of the 20th century has its own interesting features. Being proposed as a classical field theory, it can also be interpreted in the frameworks of quantum mechanics and string theory. The Kaluza-Klein theory is a five-dimensional theory and includes gravitational, electromagnetic, and scalar fields. The effects of a compact fifth dimension have been searched for in experiments at the Large Hadron Collider~\cite{CMS11} without any success up to now. Astrophysical tests of the Kaluza-Klein theory have been performed using analysis of the equations of motion and its application to galactic motion/rotation curves~\cite{Wesson95}. Particularly, the shadow of one of the black hole solutions of the theory has been studied in~\cite{Amarilla13}. On the other hand, future modifications and improvements of gravitational wave observations may give some constraints on the theory~\cite{Cardoso19,Andriot17,LIGO16b}. In \cite{Azreg20,Zhu20}, the authors have studied the effects of the Kaluza-Klein theory on the precession of a gyroscope and possible tests using X-ray spectroscopy. Here we plan to extend our previous studies by considering the shadow cast by a black hole within the Kaluza-Klein theory. 

Several solutions have been obtained within the Kaluza-Klein theory describing compact objects and particularly black holes~\cite{Horowitz11}. In Refs.~\cite{Dobiasch82,Chodos82,Gibbons86} one may find different spherically-symmetric solutions of the theory. The axially-symmetric solutions of the Kaluza-Klein theory have been obtained in Refs.~\cite{Larsen00,Rasheed95,Matos97} for the cases of four- and five-dimensional spacetimes. The black hole solution with squashed horizon has been studied in~\cite{Ishihara06,Wang06}. The solutions for black holes in higher dimensional spacetimes have been obtained in~\cite{Park98}.  


{In this paper, we study the optical properties of rotating black holes in Kaluza-Klein theory described by the four parameters of total mass, spin, electric and magnetic charges. The paper is organized as follows: The spacetime metric of a Kaluza-Klein black hole is presented and its properties are described in Sect.~\ref{metricspace}. Sect.~\ref{rtcs} is devoted to the development of the ray-tracing code for studying the photon motion and obtaining the black hole shadow. In Sect.~\ref{shadow}, the shadow of the Kaluza-Klein black hole is explored and, in addition, the light deflection angle for gravitational lensing by the black hole is studied. Finally, we summarize our main results in Sect.~\ref{Summary}.} 

\section{Spacetime metric\label{metricspace}}

For the spacetime metric, the notation developed in~\cite{Azreg20} will be used. In order to maintain consistency with our geometrized system of units, we adopt the convention of setting the speed of light, $c_0$, to unity, i.e., $c_0=1$.   Additionally, the covariant derivatives represented by $\nabla_\alpha$ and the coordinate systems $x^\alpha$ are used within the framework of the general metric ansatz for five-dimensional space-time. This ansatz encapsulates the spatial dimensions, $x^i$, $x^4$ aligns with the time coordinate, $t$, and $x^5$ designates the fifth or extra dimension, $\psi$. There are three fields involved in the simplest Kaluza-Klein theory: gravity, the dilaton, and the gauge field. The canonical form of the action in the Einstein frame can be written as follows:

\begin{equation}\label{action}
\begin{aligned}
S=\int\sqrt{-g}(\, \frac{R}{\kappa}& + \frac{1}{4}e^{\sqrt{3} \kappa\sigma } F_{\alpha\beta} F^{\alpha\beta} \\
& + \frac{1}{2} \nabla^{\alpha}\sigma\nabla_{\alpha}\sigma )\,\text{d}^4x,
\end{aligned}
\end{equation}

where $\sqrt{-g}$ represents the determinant of the four-dimensional metric tensor, $R$ is the Ricci scalar, $\sigma$ is a dilaton scalar field, $F_{\alpha\beta}$ is the gauge field, $\kappa=\sqrt{16\pi G}$ and $G$ is the gravitational constant. The metric in the Kaluza-Klein theory is given as

\begin{widetext}
\begin{equation}
ds^2=\frac{H_2}{H_1}(\text{d}\psi + {\Psi})^2 - \frac{H_3}{H_2}(\text{d}t + {T})^2 + H_1 (\frac{ \text{d} r^2}{\Delta} + \text{d}\theta^2 + \frac{\Delta}{H_3}sin^2\theta \text{d}\phi^2),
\end{equation}

where

\begin{equation}
\begin{aligned}
H_1 & = r^2 + a^2\cos^2\theta + r(p-2m)+\frac{p(p-2m)(q-2m)}{2(p+q)}-\frac{p}{2m(p+q)}\sqrt{(q^2-4m^2)(p^2-4m^2)}a\cos\theta, \\
H_2 & = r^2 + a^2\cos^2\theta + r(q-2m)+\frac{q(p-2m)(q-2m)}{2(p+q)}-\frac{q}{2m(p+q)}\sqrt{(q^2-4m^2)(p^2-4m^2)}a\cos\theta, \\
H_3 & = r^2 + a^2\cos^2\theta - 2mr, \ \ \ \ \Delta=r^2 + a^2 - 2mr, \\
\end{aligned}
\end{equation}

and the one-forms are given by

\begin{equation}
\begin{aligned}
{\Psi}=& -\frac{1}{H_2} \Big[ 2Q (r+\frac{p-2m}{2}) + \sqrt{\frac{q^3(p^2-4m^2)}{4m^2(p+q)}} a\cos\theta  \Big]dt - \frac{1}{H_2}\Big[ 2P(H_2 + a^2 \sin^2\theta)\cos\theta + \sqrt{\frac{(p^2-4m^2)}{4m^2(p+q)^3}} \\
  & \times[(p+q) (pr - m(p-2m)) + q(p^2-4m^2)]a\sin^2\theta \Big]d\phi, \\
{T}=& \frac{ (pq+4m^2)r - m(p-2m)(q-2m) }{ 2m(p+q)H_3 } \sqrt{pq} \sin^2\theta d\phi. \\
\end{aligned}
\end{equation}

The physical mass $M$, angular momentum $J$, electric charge $Q$, and magnetic charge $P$ are included in the solution via four free parameters, $m$, $a$, $p$, and $q$, respectively. The relations are written as:

\begin{equation}\label{eqM}
M=\frac{p+q}{4}, \ \ J=\frac{\sqrt{pq}(pq+4m^2)}{4m(p+q)}a, \ \ Q^2=\frac{q(q^2-4m^2)}{4(p+q)}, \ \ P^2=\frac{p(p^2-4m^2)}{4(p+q)}. 
\end{equation}

The four dimensional version of the metric in the Einstein frame is given by

\begin{equation}
d\bar{s}^2 = - \frac{H_3}{\rho^2} \text{d}t^2 - 2\frac{H_4}{\rho^2}\text{d}t \text{d}\phi + \frac{\rho^2}{\Delta}\text{d}r^2 + \rho^2 \text{d}\theta^2 + \left( \frac{-H^2_4+\rho^4\Delta\sin^2\theta}{\rho^2H_3} \right)\text{d}\phi^2
\end{equation}
where
\begin{equation}
\begin{aligned}
\frac{H_1}{M^2} &= \frac{8(b-2)(c-2)b}{(b+c)^3} + \frac{4(b-2)x}{b+c} + x^2 - \frac{2b\sqrt{(b^2-4)(c^2-4)}\alpha\cos\theta }{(b+c)^2} + \alpha^2\cos^2\theta, \\
\frac{H_2}{M^2} &= \frac{8(b-2)(c-2)c}{(b+c)^3} + \frac{4(c-2)x}{b+c} + x^2 - \frac{2c\sqrt{(b^2-4)(c^2-4)}\alpha\cos\theta }{(b+c)^2} + \alpha^2\cos^2\theta, \\ 
\frac{H_3}{M^2} &= x^2 + \alpha^2\cos^2\theta - \frac{8x}{b+c}, \ \ \ \frac{\Delta}{M^2}=x^2+\alpha^2-\frac{8x}{b+c}, \\
\frac{H_4}{M^3} &= \frac{ 2\sqrt{bc} \left[ (bc+4) (b+c)x - 4(b-c)(c-2) \right]\alpha\sin^2\theta }{ (b+c)^3 },
\end{aligned}
\end{equation}
\end{widetext}
and $\rho=\sqrt{H_1H_2}$. Here $\alpha$, $b$, $c$, and $x$ are the dimensionless parameters that are defined as $\alpha\equiv a/M$, $b\equiv p/m$, $c\equiv q/m$, and $x\equiv r/M$. From Eq.~\ref{eqM}, the free parameter $m$ can be related to the physical mass $M$ as
\begin{equation}
m=\frac{4M}{b+c}.
\end{equation}
The Kaluza-Klein metric reduces to the Kerr metric when the electric and magnetic charges vanish, and the spin parameter $\alpha$ equals the Kerr solution's dimensionless spin $a_{*}$.
The spacetime has two horizons
\begin{equation}
r_{\pm}=m\pm\sqrt{m^2-a^2}, \label{eqhorizon}
\end{equation}
which are obtained by solving $\Delta=0$. In terms of the dimensionless parameters Eq.~\ref{eqhorizon} becomes
\begin{equation}
x_{\pm}=\frac{4 \pm \sqrt{16-\alpha^2 (b+c)^2 } }{b+c}.\label{eqhorizon2}
\end{equation}
The Reissner-Nordstrom solution of general relativity is obtained by setting $b=c$ and $\alpha=0$. We recover the Kerr solution when $b=c=2$. However, the Kerr-Newman solution is not recovered when the magnetic charge vanishes. For our analysis, we consider the case when $b=c$, which is a black hole with electric and magnetic charges. Note that this case can only be considered as a toy-model, since macroscopic black holes have a negligible electric charge.

Next, we make sure that the spacetime does not possess any pathologies. The bounds on the free parameters are obtained by preserving the metric structure everywhere outside the horizon. First, from the definitions of $Q^2$ and $P^2$ in Eq.~\ref{eqM}, we get the following constraints: $q\geq2m$, $p\geq2m$, or $b\geq2$, $c\geq2$. Plugging these into Eq.~\ref{eqhorizon2}, we come to the bound on $\alpha$:
\begin{equation}
\alpha^2\leq1 \ \ \text{or} \ \ -1<\alpha<1. \label{eq:alpharange}
\end{equation}
The upper bound of $b$ ($b=c$) can be obtained by using Eqs.~\ref{eqhorizon2}-\ref{eq:alpharange} as
\begin{equation}
2\leq b \leq \frac{2}{|\alpha|}. \label{eq:brange}
\end{equation}

\section{Ray-tracing code for photons\label{rtcs}}

The photon sphere of a black hole spacetime separates geodesics falling into the event horizon from those escaping to spatial infinity~\cite{Claudel01}. We observe the photon sphere in the form of the black hole shadow. In the case of generic spacetimes, we cannot solve the null-geodesic equations analytically and we have to solve the photon motion numerically in order to calculate the shadow. In the Kerr spacetime, the photon sphere is described by an analytical solution that is used to calculate the shadow. 
The Kaluza-Klein black hole metric we are studying is not separable and, therefore, the shadow will be calculated numerically by applying a general relativistic ray-tracing code, which will be described below. 

The ray-tracing code we apply calculates photon trajectories near the black hole and is based on the code used in~\cite{Ayzenberg18, Gott19, Abdikamalov19}, which was first described in~\cite{Psaltis12}. To obtain the equations that represent $t$- and $\phi$- component of the photon position, we exploit the property of all stationary and axisymmetric spacetimes, namely, that a test particle's four momentum is related to the specific energy $E$ and the $z$-component of the specific angular momentum $L_z$ as $p_t =-E$ and $p_\phi = L_z$. These relations lead to two first-order differential equations 

\begin{equation}\label{t_comp}
\begin{aligned}
&\frac{dt}{d\lambda'} =& -\frac{Bg_{t\phi} + g_{\phi\phi}}{g_{tt}g_{\phi\phi}-g_{t\phi}^{2}}, 
\end{aligned}
\end{equation}
\begin{equation}\label{phi_comp}
\begin{aligned}
&\frac{d\phi}{d\lambda'} =& \frac{g_{t\phi}+Bg_{tt}}{g_{tt}g_{\phi\phi}-g_{t\phi}^{2}}, 
\end{aligned}
\end{equation}

where $\lambda'\equiv E / \lambda $ denotes the normalized affine parameter and $B\equiv  L_z/E$ is the impact parameter.

The $r$- and $\theta$-components of the photon position are obtained by solving the second-order geodesic equations for a generic axisymmetric metric
\begin{equation}\label{geod1}
	\begin{aligned}
		& \frac{d^2r}{d\lambda'^2}=
		-\Gamma_{tt}^{r}\left(\frac{dt}{d\lambda'}\right)^2
		-\Gamma_{rr}^{r}\left(\frac{dr}{d\lambda'}\right)^2
		-\Gamma_{\theta\theta}^{r}\left(\frac{d\theta}{d\lambda'}\right)^2 \\
		&  \ \ \ \ \ \ \ \ \ \ -\Gamma_{\phi\phi}^{r}\left(\frac{d\phi}{d\lambda'}\right)^2
		-2\Gamma_{t\phi}^{r}\left(\frac{dt}{d\lambda'}\right)\left(\frac{d\phi}{d			\lambda'}\right) \\
		& \ \ \ \ \ \ \ \ \ \ -2\Gamma_{r\theta}^{r}\left(\frac{dr}{d\lambda'}\right)			\left(\frac{d	\theta}{d\lambda'}\right), \\
			\end{aligned}
\end{equation}
\begin{equation}\label{geod2}
	\begin{aligned}
		& \frac{d^2\theta}{d\lambda'^2}=
		-\Gamma_{tt}^{\theta}\left(\frac{dt}{d\lambda'}\right)^2
		-\Gamma_{rr}^{\theta}\left(\frac{dr}{d\lambda'}\right)^2 - 			\Gamma_{\theta\theta}^{\theta}\left(\frac{d\theta}{d\lambda'}\right)^2 \\
		&  \ \ \ \ \ \ \ \ \ \ -\Gamma_{\phi\phi}^{\theta}\left(\frac{d\phi}{d\lambda'}\right)^2 
		-2\Gamma_{t\phi}^{\theta}\left(\frac{dt}{d\lambda'}\right)\left(\frac{d			\phi}{d\lambda'}\right) \\
		& \ \ \ \ \ \ \ \ \ \  -2\Gamma_{r\theta}^{\theta}\left(\frac{dr}{d				\lambda'}\right)	\left(\frac{d\theta}{d\lambda'}\right),
	\end{aligned}
\end{equation}
where the Christoffel symbols of the metric are given by $\Gamma_{\mu\nu}^{\sigma}$. 

We set the reference frame and coordinate system in such a way that the black hole is at the origin and $z$-axis coincides with the axis of the spin angular momentum. In the code we set the physical mass as $M=1$, since the shape of the black hole shadow does not depend on the physical mass $M$, which only affects the size of the shadow. In the calculations, the distance, azimuthal angle, and polar angle of the observing screen are set to $d=1000$, $\theta = \iota$, and $\phi=0$, respectively. The relations $\bar{\alpha}=r_{scr}\cos(\phi_{scr})$ and $\bar{\beta}=r_{scr}\sin(\phi_{scr})$ link the polar coordinates $r_{scr}$ and $\phi_{scr}$ used on the screen and the celestial coordinates $(\bar{\alpha}, \bar{\beta})$ on the observer’s sky.

The system of equations (\ref{geod1})-(\ref{geod2}) are evolved backwards in time, since only the final positions and momenta of the photons are known. Each photon originates with some initial coordinates on the screen having a four-momentum perpendicular to the screen. The latter condition is necessary for modeling the position of the screen located very far from the source, since photons hitting the screen at spatial infinity would move perpendicular to the screen at a distance of $d$.

Each photon's initial position and four-momentum in the Boyer-Lindquist coordinates of the black hole spacetime are expressed as~\cite{Bambi17e}
\begin{equation}\label{in_pos_r}
	\begin{aligned}
		& r_i =\left(d^2+\bar{\alpha}^2+\bar{\beta}^2\right)^{1/2}      ,\\
	\end{aligned}
\end{equation}
\begin{equation}\label{in_pos_theta}
	\begin{aligned}
		& \theta_i=\arccos\left(\frac{d\cos\iota+\bar{\beta}\sin\iota}{r_i}\right),\\
	\end{aligned}
\end{equation}
\begin{equation}\label{in_pos_phi}
	\begin{aligned}
		& \phi_i = \arctan\left(\frac{\bar{\alpha}}{d\sin\iota-\bar{\beta}\cos\iota}\right),
	\end{aligned}
\end{equation}
and
\begin{equation}\label{in_momen_r}
	\begin{aligned}
	& \left(\frac{dr}{d\lambda'}\right)_i=\frac{d}{r_i},\\
	\end{aligned}
\end{equation}

\begin{equation}\label{in_momen_theta}
	\begin{aligned}
	& \left(\frac{d\theta}{d\lambda'}\right)_i=\frac{-\cos\iota+\frac{d}{r_i^2}(d\cos\iota+\bar{\beta}\sin\iota)}{\sqrt{r_i^2-(d\cos\iota+\bar{\beta}\sin\iota)^2}},\\
	\end{aligned}
\end{equation}
\begin{equation}\label{in_momen_phi}
	\begin{aligned}
	& \left(\frac{d\phi}{d\lambda'}\right)_i=\frac{-\bar{\alpha}\sin\iota}{\bar{\alpha}^2+({d\sin\iota-\bar{\beta}\cos\iota})^2},\\
	\end{aligned}
\end{equation}

\begin{widetext}
\begin{equation}\label{in_momen_t}
    \begin{aligned}
        \left(\frac{d t}{d \lambda^{\prime}}\right)_i=  -\frac{g_{t \phi}}{g_{t t}}\left(\frac{d \phi}{d \lambda^{\prime}}\right)_i
        +\sqrt{\frac{g_{t \phi}^2}{g_{t t}^2}\left(\frac{d \phi}{d \lambda^{\prime}}\right)_i^2-\left[\frac{g_{r r}}{g_{t t}}\left(\frac{d r}{d \lambda^{\prime}}\right)_i^2+\frac{g_{\theta \theta}}{g_{t t}}\left(\frac{d \theta}{d \lambda^{\prime}}\right)_i^2+\frac{g_{\phi \phi}}{g_{t t}}\left(\frac{d \phi}{d \lambda^{\prime}}\right)_i^2\right].}
    \end{aligned}
\end{equation}
\end{widetext}

We can find the component $(dt/d\lambda')_i$ from the fact that the norm of the photon four-momentum is zero. The conserved quantity $B$ required in Eqs. (\ref{t_comp}) and (\ref{phi_comp}) can be calculated from the initial conditions.

\begin{figure*}[ht!]
\centering
\includegraphics[width=1.0\linewidth]{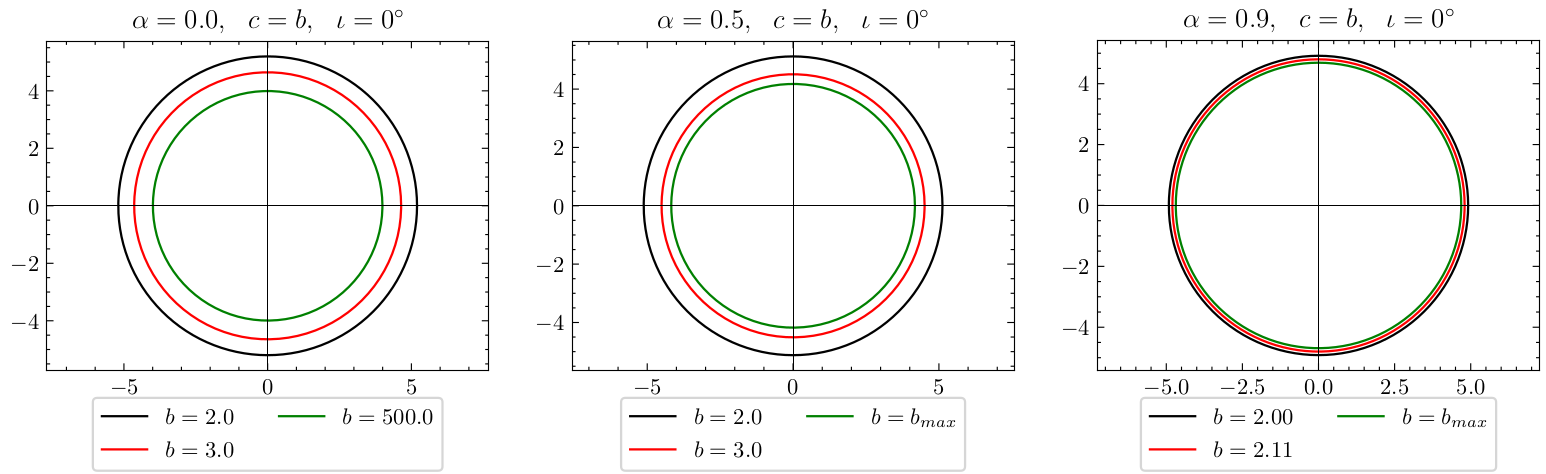}
\caption{Shadow images calculated using the ray-tracing code at an inclination angle $\iota=0^{\circ}$. {Left panel: shadow silhouettes for $\alpha=0.0$ with $b=[2, 3, 500]$. Central panel: shadow silhouettes for $\alpha=0.5$ with $b=[2, 3, b_{max}]$. Right panel: shadow silhouettes for $\alpha=0.9$ with $b=[2.0, 2.11, b_{max}]$. $b_{max}=\frac{2}{\alpha}$.} As expected, these shadow images are circles, and only their size is affected by $\alpha$ and $b$.}\label{f_shadow1}
\end{figure*}

\begin{figure*}[ht!]
\centering
\includegraphics[width=1.0\linewidth]{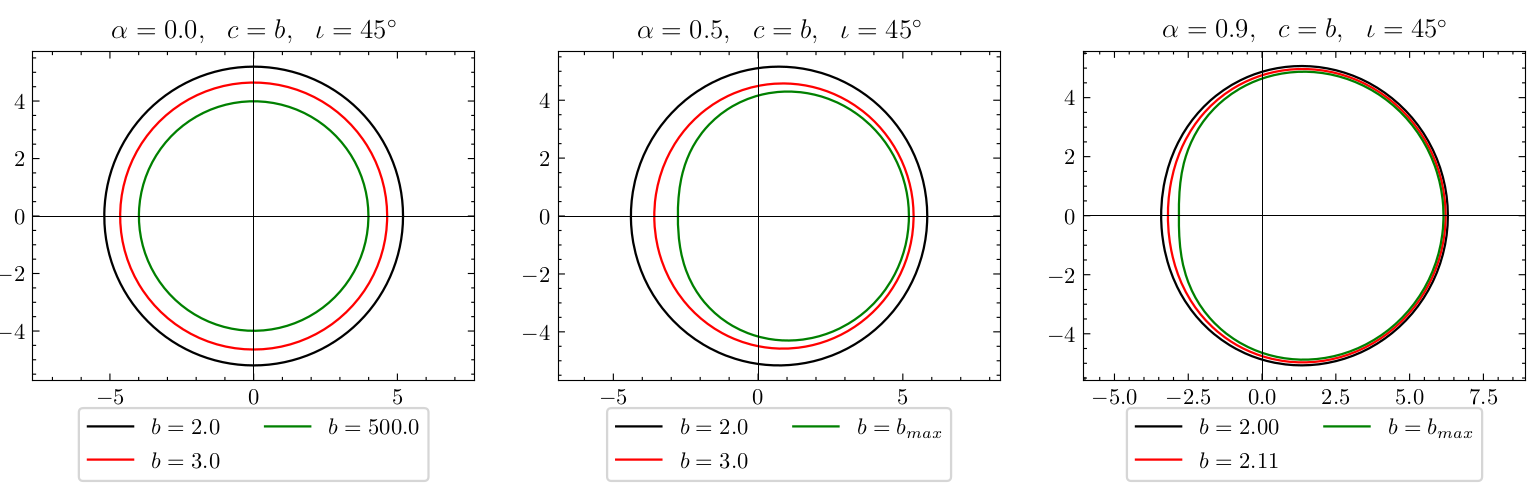}
\caption{Shadow images calculated using the ray-tracing code at an inclination angle $\iota=45^{\circ}$. {Left panel: shadow silhouettes for $\alpha=0.0$ with $b=[2, 3, 500]$. Central panel: shadow silhouettes for $\alpha=0.5$ with $b=[2, 3, b_{max}]$. Right panel: shadow silhouettes for $\alpha=0.9$ with $b=[2.0, 2.11, b_{max}]$. $b_{max}=\frac{2}{\alpha}$.} It is easy to see that as the spin parameter increases the barycenter of the images moves away from the center and the distortion of the shadow from a circle increases. Also, same effects are achieved by increasing $b$ for fixed $\alpha$}\label{f_shadow2}
\end{figure*}

\begin{figure*}[ht!]
\centering
\includegraphics[width=1.0\linewidth]{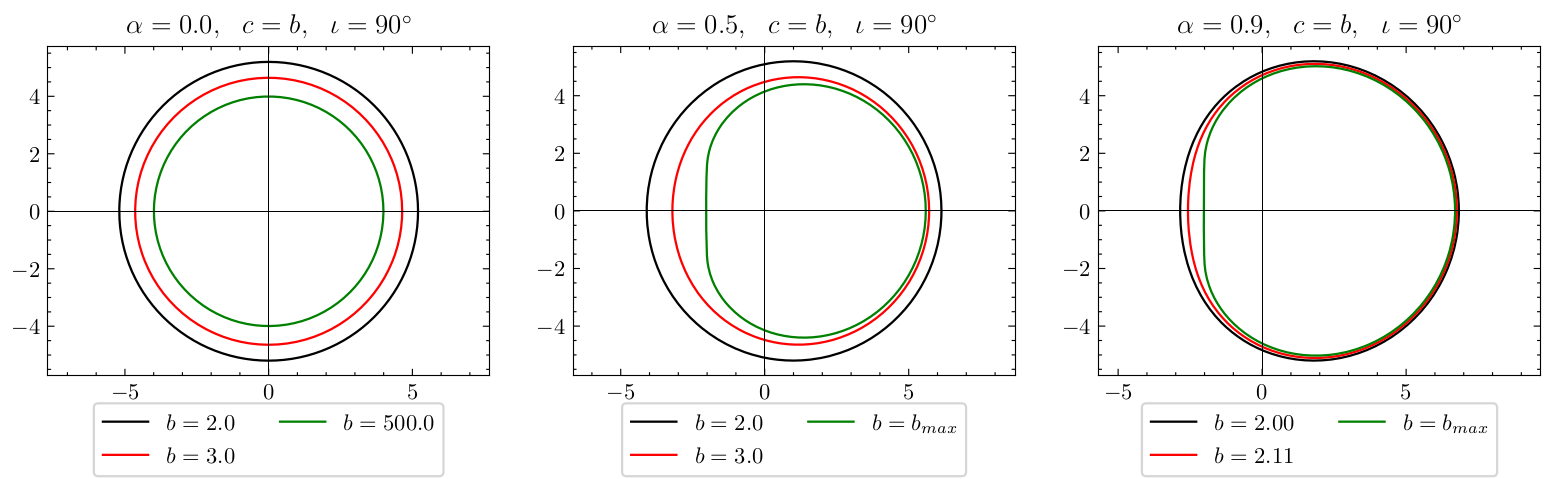}
\caption{Shadow images calculated using the ray-tracing code at an inclination angle $\iota=90^{\circ}$.  {Left panel: shadow silhouettes for $\alpha=0.0$ with $b=[2, 3, 500]$. Central panel: shadow silhouettes for $\alpha=0.5$ with $b=[2, 3, b_{max}]$. Right panel: shadow silhouettes for $\alpha=0.9$ with $b=[2.0, 2.11, b_{max}]$. $b_{max}=\frac{2}{\alpha}$.}  Similar to the case with $\iota=45^{\circ}$, increasing the spin causes horizontal displacement and distortion from a circle. Same effects are achieved by increasing $b$ for fixed $\alpha$}\label{f_shadow3}
\end{figure*}

The initial conditions on the screen are sampled by the code in the following way. The location of the shadow boundary separating photons falling into the horizon from those fleeing into spatial infinity is searched in the range $0 \leq r_{scr} \leq 20$ for each $\phi_{scr}$ in the range $0 \leq \phi_{scr} \leq 2\pi$ and in increments $\pi / 180$. Photons intersecting $r=r_{hor} + \delta r$, with $\delta r = 10^{-6}$, are considered captured in the horizon, while all other photons reach $r > d = 1000$, hence escaping to spatial infinity. The accurate value of $r_{scr}$ that represents the shadow boundary for current $\phi_{scr}$ is determined by zooming in the boundary to an accuracy $\delta r_{scr} \thicksim 10^{-3}$. This methodology can accurately calculate the black hole shadow more efficiently than fine sampling the entire screen. Fig.~\ref{f_shadow1} shows the shadow silhouettes calculated using the ray-tracing code for $\alpha=[0.0, 0.5, 0.9]$ and several values of the deformation parameter $b$ at an inclination angle $\iota=0^{\circ}$. Fig.~\ref{f_shadow2} and Fig.~\ref{f_shadow3} illustrate the same images calculated at inclination angles $\iota=45^{\circ}$ and $\iota=90^{\circ}$. When $\iota=0^{\circ}$, the shadow images are circles with different sizes affected by $\alpha$ and $b$. For $\alpha\neq0$, the barycenter of the images moves away from the center and the distortion of the shadow from a circle increases as spin increases. Also, the same effects are achieved by increasing $b$ for fixed $\alpha$.

\section{Shadow of Kaluza-Klein black holes \label{shadow} }

In this section, we will study the apparent shape of the Kaluza-Klein black hole shadow. 
We consider the celestial coordinates $\bar{\alpha}$ and $\bar{\beta}$~(see \cite{Vazquez04,Abdujabbarov16b} for reference) defined as,
\begin{eqnarray}
\bar{\alpha} &=&\underset{r_0 \rightarrow \infty}{\lim}\left(-r_0^2 \sin \iota \frac{d\phi}{dr}\right) \ , \label{eq:14}\\
\bar{\beta} &=&\underset{r_0 \rightarrow \infty}{\lim}\left(r_0^2 \frac{d\theta}{dr}\right) \ , \label{eq:15}
\end{eqnarray}
where $r_0$ denotes the distance between the observer and black hole and $\iota$ is the inclination angle between the black hole spin axis and the observer's line of sight. 
We will apply the coordinate independent formalism used in~\cite{Ayzenberg18} to describe the relationship between the shape of the shadow and the deformation parameter. The horizontal displacement from the center of the image $D$, the average radius of the sphere $\langle R \rangle$, and the asymmetry parameter $A$ parametrize the shape of the shadow. We will get similar results with any other chosen parametrizations (see e.g. \cite{Tsukamoto14,Abdujabbarov15}). The horizontal displacement $D$ registers the shift of the shadow center from the center of the black hole, and it is defined as
\begin{eqnarray}
D \equiv \frac{|\bar{\alpha}_{max}-\bar{\alpha}_{min}|}{2},
\end{eqnarray}
where $\bar{\alpha}_{max}$ and $\bar{\alpha}_{min}$ represent the maximum and 
\begin{figure*}[ht!]
\centering
\includegraphics[width=1.0\linewidth]{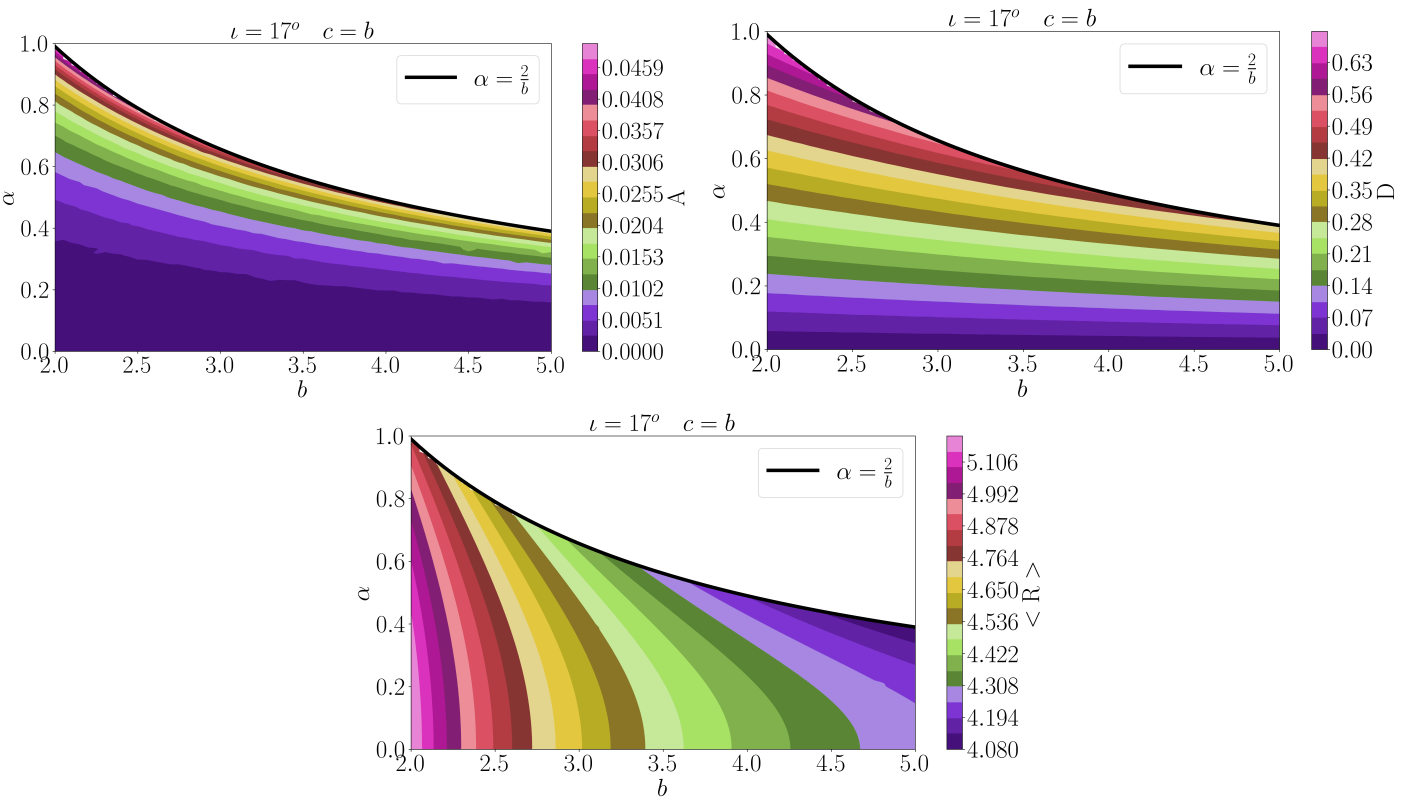}
\caption{Density plot of the shadow parametrizations; asymmetry parameter $A$ (top left panel), horizontal displacement $D$ (top right panel) and average radius $\langle R \rangle$ (bottom panel). The black solid line at $\alpha=2/b$ marks the upper boundary of the parameter $b$. The Kerr case is recovered for $b=2$, the left edge of the plot.}\label{f_parametrization}
\end{figure*}
\begin{figure*}[ht!]
\centering
\includegraphics[width=1.0\linewidth]{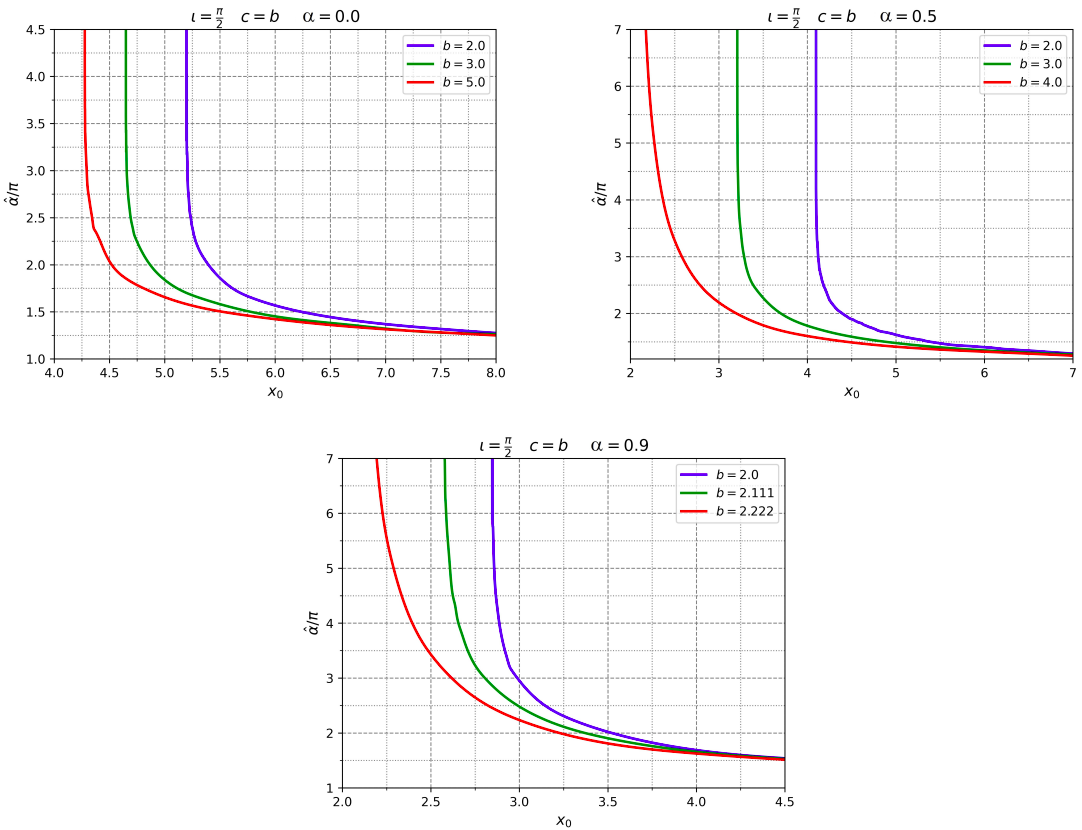}
\caption{The deflection angle ($\hat{\alpha}$) of photons with impact parameter $x_0=B$ for $\alpha=0.0$ (top left panel), $\alpha=0.5$ (top right panel) and $\alpha=0.9$ (bottom panel). $b=2$ corresponds to the Kerr case. }\label{f_def_angle}
\end{figure*}
minimum horizontal coordinates of the image on the observing screen, respectively. We can safely ignore the vertical displacement of the image, since the spacetime in which we are working is axisymmetric. The average radius $\langle R \rangle$ defines the average distance between the boundary and the center of the shadow, which is defined by
\begin{eqnarray}\label{Rav}
\langle R \rangle \equiv \frac{1}{2\pi} \int_{0}^{2\pi} R(\vartheta)d\vartheta,
\end{eqnarray}
where $R(\vartheta) \equiv \left[(\bar{\alpha}-D)^2+\bar{\beta}(\bar{\alpha})^2\right]^{1/2}$ and $ \vartheta \equiv \tan^{-1}[\bar{\beta}(\bar{\alpha})/\bar{\alpha})]$. The asymmetry parameter $A$ characterizes the distortion of the shadow from a circle, and it is defined by
\begin{eqnarray}\label{Asym}
A \equiv 2\left[\frac{1}{2\pi}\int_{0}^{2\pi} \left(R-\langle R \rangle \right)^2 d\vartheta \right]^{1/2}.
\end{eqnarray}

Fig.~\ref{f_parametrization} shows the density plot of the shadow parametrizations: asymmetry parameter $A$ (top left panel), horizontal displacement $D$ (top right panel) and average radius $\langle R \rangle$ (bottom panel). The black solid curves mark the upper boundary of the parameter $b$ ($\alpha=2/b$). By varying $b$, the same values of the asymmetry parameter $A$ and the horizontal displacement $D$ can be obtained for different values of $\alpha$. As we can see, $\alpha$ is the main parameter that affects the values of $D$ and $A$, and $b$ is the main parameter that affects $<R>$. Additionally,  the influence of the parameter $\alpha$ on both $A$ and $D$ manifests distinct behaviors. In the former case, $\alpha$ exhibits a power-law relationship, while in the latter case, the impact is approximately linear.

\subsection{Deflection angle}\label{sec_def_angle}

\begin{figure*}[ht!]
\includegraphics[width=1.0\linewidth]{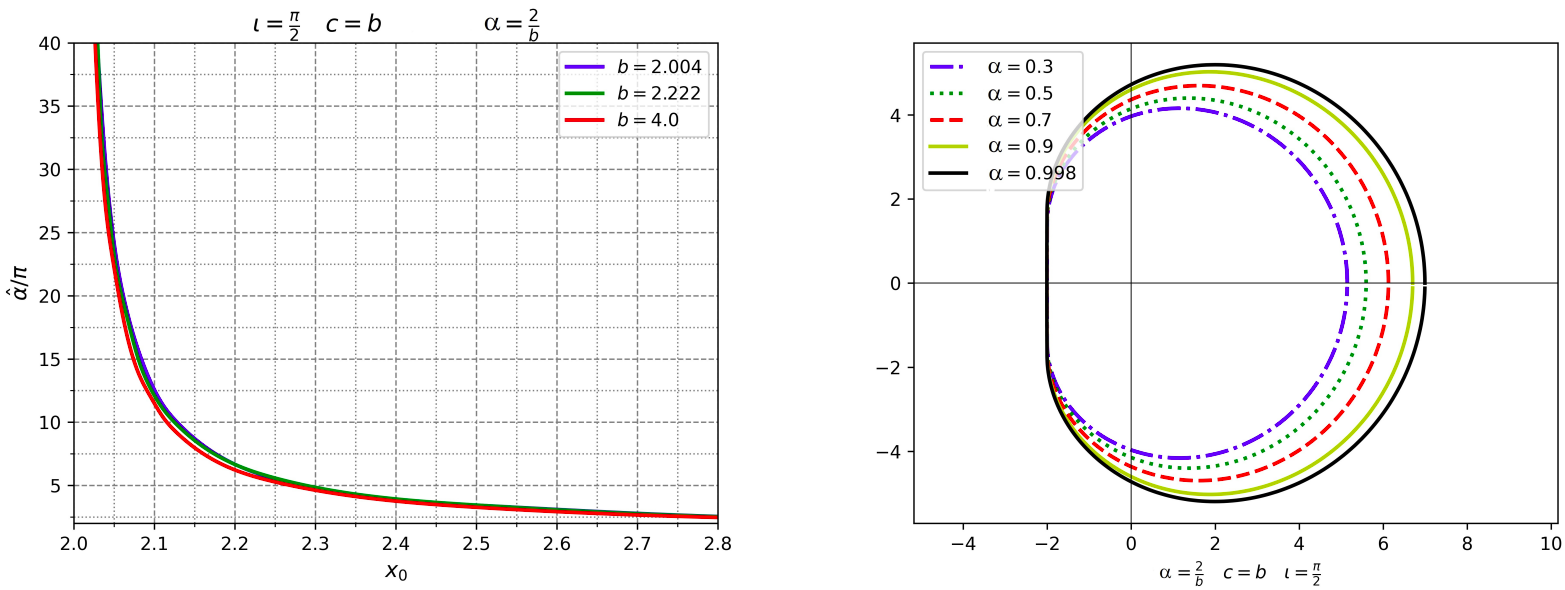}
\caption{Deflection angles and shadow images for several values of the spin parameter $\alpha$ with maximum allowed $b$. We can see that the deflection angles are nearly same. Also, from the corresponding shadow images, the horizontal displacement and the distortions take similar values.}\label{f_d_s}
\end{figure*}

Here we calculate the deflection angle of photons that pass near the black hole using the modified version of the ray-tracing code described above. The calculation algorithms are similar to those discussed in Sec.~\ref{rtcs}. We restrict the entire photon trajectories to the equatorial plane of the Kaluza-Klein black hole. We initialize photons on the screen with certain celestial coordinates $(\bar{\alpha}, \bar{\beta})$. Setting $\bar{\alpha}\neq0$, $\bar{\beta}=0$ and $\iota=\pi/2$ in Eqs.~\ref{in_pos_r}-\ref{in_momen_t} lead to $\theta=\pi/2$, $\dot{\theta}=0$ with non-zero $\dot{r}$, $\dot{\phi}$, so we obtain the photon trajectories lying in the equatorial plane. $\bar{\alpha}$ is chosen in such a way that photons with the impact parameter $B$ approach the photon ring of the black hole at the minimum coordinate difference $d=10^{-7}M$ without crossing it. The deflection angle from a straight line is calculated by capturing the initial and final positions of the photon. Fig.~\ref{f_def_angle} illustrates the calculated values of the deflection angle $\hat{\alpha}/\pi$ as a function of photons with impact parameter $B$ for $\alpha=[0.0, 0.5, 0.9]$. Increasing $b$ shifts the impact parameter $x_0=B$ towards lower values. This can be explained by an increase of the horizontal displacement and a distortion from a circle. Fig.~\ref{f_d_s} shows the deflection angles and shadow images for several values of the spin parameter $\alpha$ with maximum allowed $b$. We can see that the deflection angles are nearly the same. Also, from the corresponding shadow images, the horizontal displacement and the distortion take similar values.

\subsection{A simple constraint of the Kaluza-Klein black hole parameters using the M87* results}

{Now we plan to study whether one can place constraints on the parameters of a Kaluza-Klein BH using the M87* black hole shadow observations. We follow the precept of \cite{EHT19a} that  EHT observations are related to the shadow of a rotating Kerr BH. Based on the results from the shadow of M87* BH observations, we constrain the parameter $b$ (assuming $c=b$) and the spin of the Kaluza-Klein black hole using the average radius (Eq.~\ref{Rav}) and the asymmetry parameter (Eq.~\ref{Asym}) of the BH shadow. Note that this approach is a simplification of the correct analysis, in which the calculated shadows should be convolved with the response of the instrument, followed by a final comparison of the output data with EHT data.

According to observations made by the EHT collaboration, the angular size of the observed shadow is $\Delta \theta_{s h}=42 \pm 3 \rm{\mu as}$, and a deviation $\delta$ from circularity is less than $10 \%$ }

$$
\delta = \frac{1}{\langle R\rangle}\left[\frac{1}{2 \pi} \int_{0}^{2 \pi}(R-\langle R\rangle)^{2} d \vartheta\right]^{1 / 2}
$$

$$
\delta = \frac{A}{2 \langle R\rangle} \leq 0.1
$$

{Furthermore, we take the standard values for the distance to $\mathrm{M} 87^{*}$ to be $D=(16.8 \pm 0.8) \mathrm{Mpc}$ and the total mass of the object to be $M=(6.5 \pm 0.7) \times 10^{9} M_{\odot}$. Taking into account the accepted values of physical quantities, the average radius of the BH shadow is evaluated as}

$$
\langle R\rangle=\frac{D \Delta \theta_{s h}}{2 M}=5.50 \pm 0.75.
$$

\begin{figure*}[ht!]
\centering
\includegraphics[width=0.8\linewidth]{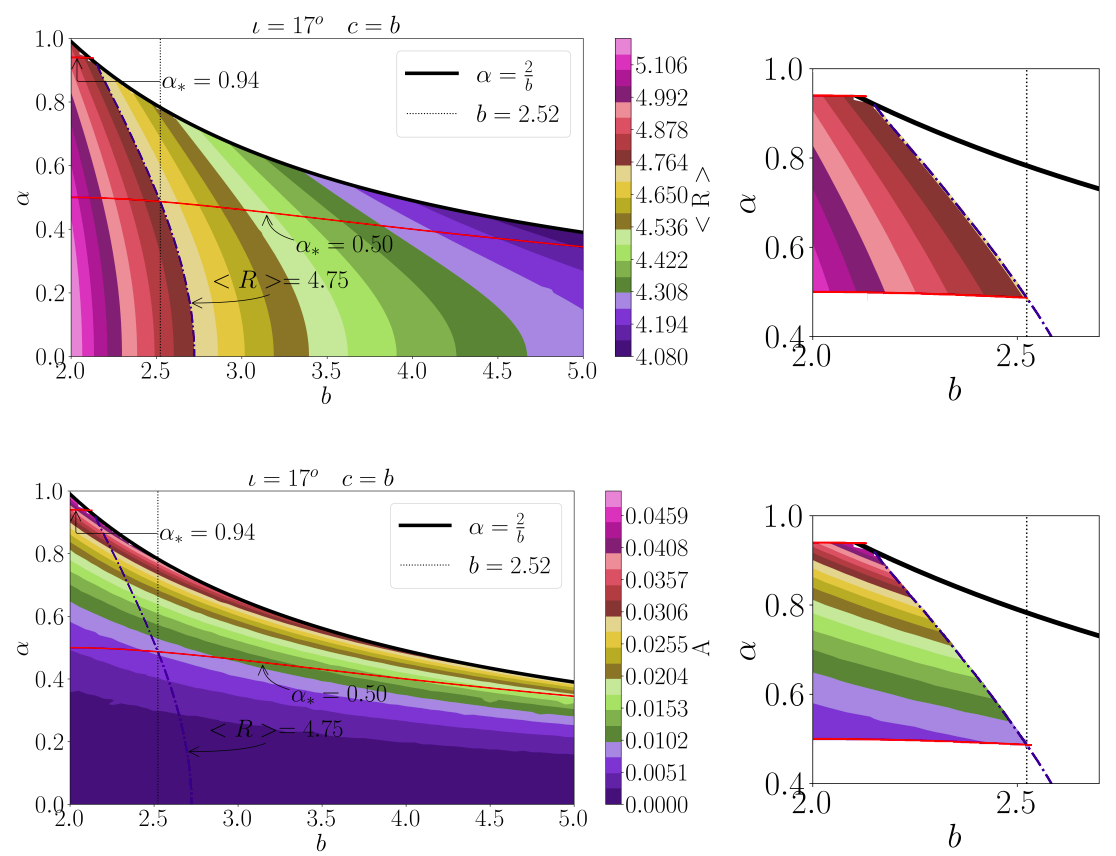}
\caption{The curves $\alpha_{*}= 0.5$ and $\alpha_{*}= 0.94$ (red lines) as well as the line $\langle R\rangle = 4.75$ are plotted for given BH spin range of $0.5 \leq \alpha_{*} \leq 0.94$ . A constraint for the parameter $b<b_0$, where $b_0 = 2.52$ is obtained according to which the asymmetry  parameter $A$ of the BH shadow is always less than 0.05.}\label{constrain}
\end{figure*}

 In Fig.~\ref{constrain},  the average radius and the asymmetry parameter of the BH shadow are shown for different values of the parameters $\alpha$ and $b$. Here, we have taken the inclination angle to be $\iota=17^{\circ}$, which the observed M87 jet axis makes with respect to the line of sight. Moreover, according to the observations by EHT collaboration, the BH spin lies within the range $0.5 \leq \alpha_{*} \leq 0.94$, where $\alpha_{*}=J / M^2$ (here $J$ is the angular momentum of BH).

The maximum value for the BH shadow radius is $6.25$. For the Kaluza-Klein black hole, the greatest value is the average radius of the shadow $\approx5.20$, so we have a limit on the size of the BH shadow from below, i.e. as $\langle R\rangle \geq 4.75$. Given the spin range $0.5 \leq \alpha_{*} \leq 0.94$ and using Eq.~\ref{eqM}, we plot the curves $\alpha_{*}= 0.5$ and $\alpha_{*}= 0.94$ (red lines in Fig.~\ref{constrain}) as well as the line $\langle R\rangle = 4.75$. As a result, we obtain the largest possible upper bound for the parameter$b$, $b<b_0$, where $b_0 = 2.52$. With such a constraint, the asymmetry  parameter $A$ of the BH shadow is always less than 0.05.

{\subsection{Images of a Kaluza-Klein black hole surrounded with thin accretion disk}}

In this subsection we explore the Novikov-Thorne model of a thin accretion disk around a Kaluza-Klein black hole.
To obtain an image of the accretion disk around the black hole, we use the ray-tracing code developed and presented in Sect.~\ref{rtcs}. A pivotal component in the creation of the image of a thin disk involves the radiation flux emission profile. When considering a massive test particle that orbits the metric in the
Kaluza-Klein theory on a circular orbit within the equatorial plane ($\theta = \pi/2$), the effective potential for the particle takes the form: 

\begin{equation}
V_{\mathrm{eff}}(r)=-1+\frac{E^{2} g_{\phi \phi}+2 E L g_{t \phi}+L^{2} g_{t t}}{g_{t \phi}^{2}-g_{t t} g_{\phi \phi}}.
\end{equation}

The specific energy $E$  and the specific angular momentum $L$ 
can be expressed as:

\begin{equation}
E=-\frac{g_{t t}+\Omega g_{t \phi}}{\sqrt{-g_{t t}-2 \Omega g_{t \phi}-\Omega^{2} g_{\phi \phi}}},
\end{equation}
\begin{equation}
L=\frac{\Omega g_{\phi \phi}+g_{t \phi}}{\sqrt{-g_{t t}-2 \Omega g_{t \phi}-\Omega^{2} g_{\phi \phi}}},
\end{equation}

where $\Omega$ is the angular velocity
\begin{equation}
\Omega=\frac{-g_{t \phi, r} \pm \sqrt{-\left(g_{t \phi, r}^{2}+g_{\phi \phi, r} g_{t t, r}\right)}}{g_{\phi \phi, r}}.
\end{equation}

The inner edge of the thin accretion disk is determined by the innermost stable circular orbit (ISCO). The ISCO's location is derived from the following conditions:

\begin{equation}
V_{\mathrm{eff}}(r)= V_{\mathrm{eff}}^{\prime}(r)= V_{\mathrm{eff}}^{\prime \prime}(r)=0.  \label{ConISCO}
\end{equation}

\begin{figure}[ht!]
    \centering

     \includegraphics[width=0.9\linewidth]{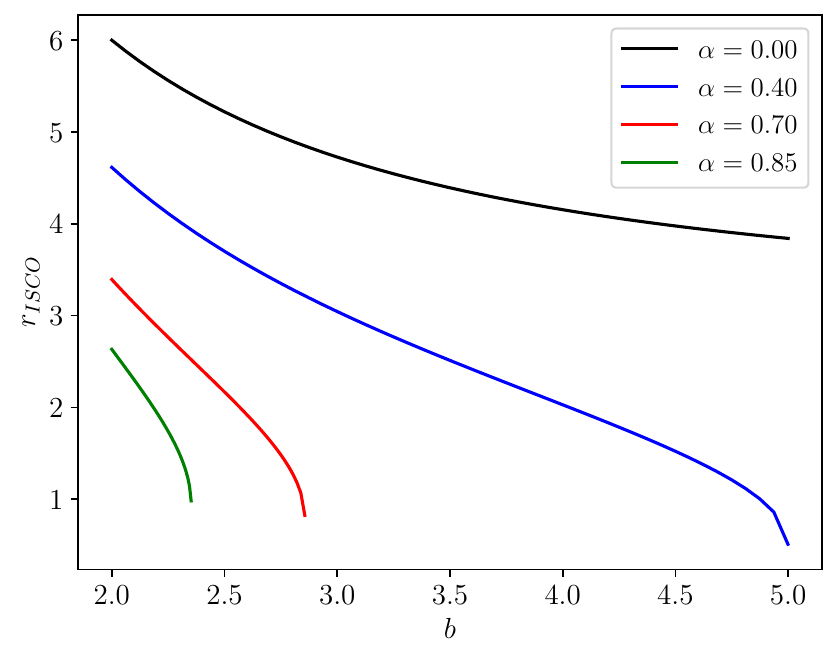}
    \caption{{The ISCO as a function of the  parameter $b$ for different values of the parameter $\alpha=[0.0,0.4,0.7,0.85]$.}} 
    \label{ISCO}
\end{figure}

Fig.~\ref{ISCO} illustrates the radius of the innermost stable circular orbit ($r_{ISCO}$)  as a function of the  parameter $b$ for different values of the parameter $\alpha$ based on the solution derived from Eq.~\ref{ConISCO}. As evident from the figure, the value of $r_{ISCO}$ decreases as the parameters $\alpha$ and $b$ increase.

By selecting  initial conditions for photons as  the photons cross the accretion disk, one can calculate the energy flux  for them as
\begin{equation}
F(r)=-\frac{\dot{M}}{4 \pi \sqrt{-g}} \frac{\Omega^{\prime}}{(E-\Omega L)^{2}} \int_{r_{\text {in }}}^{r}(E-\Omega L) L^{\prime} d r\ ,
\end{equation}
where $\dot{M}=d M / d t$ is the mass accretion rate and $r_{\text {in }}$ is the radius of the inner edge of the accretion disk, which is equivalently equal to $r_{ISCO}$.

The photon flux as detected by a distant observer is
\begin{equation}
F_{\mathrm{obs}}=g_{red}^{4} F
\end{equation}
where $g_{red}$ is the redshift factor which comprises the effects of both gravitational redshift and Doppler shift
\begin{equation}
g_{red}=\frac{k_{\mu} u_{o}^{\mu}}{k_{\nu} u_{e}^{\nu}}=\frac{\sqrt{-g_{t t}-2 g_{t \phi} \Omega-g_{\phi \phi} \Omega^{2}}}{1-B \Omega}
\end{equation}

\begin{figure*}
\centering
\includegraphics[width=0.8\linewidth]{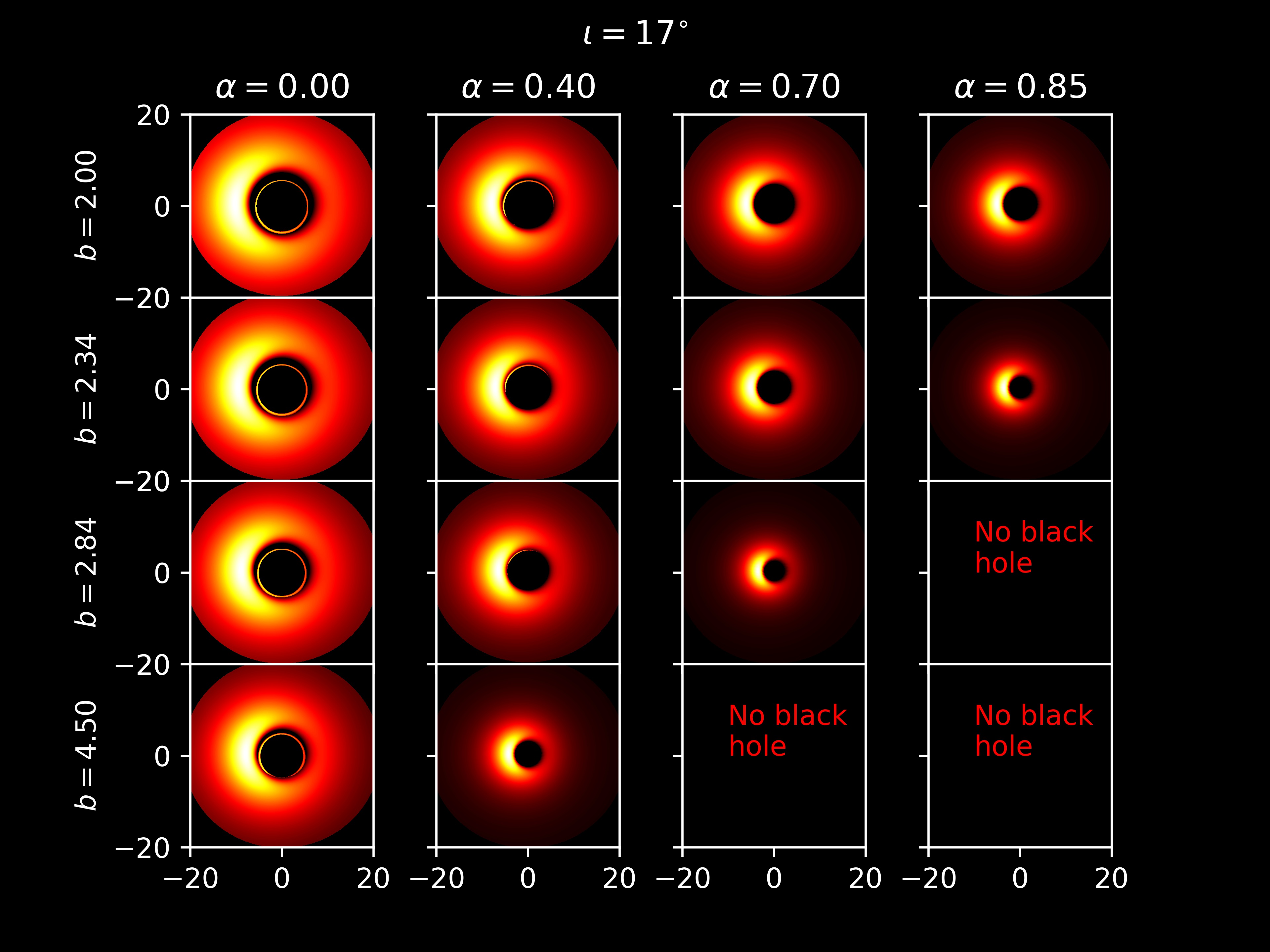}

\ \ \ \

\includegraphics[width=0.8\linewidth]{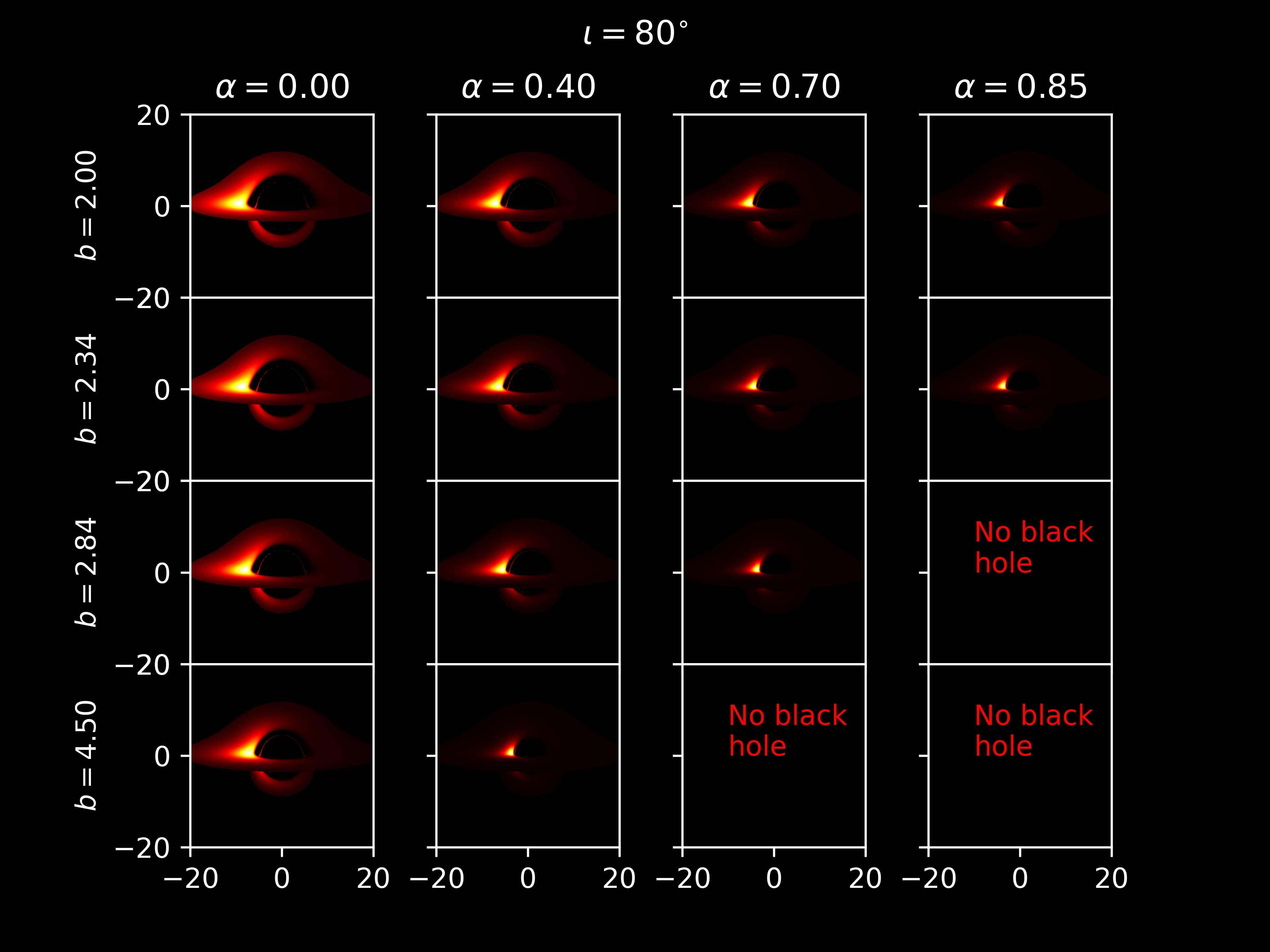}
\caption{Synthetic images of an accretion disk around a Kaluza-Klein black hole with the inclination angle $\iota = 17^{\circ}$ (top panel) and $80^{\circ}$ (bottom panel)  for different values of the BH parameters $\alpha$ and $b$.}\label{accdisk17}
\end{figure*}

Figure \ref{accdisk17} illustrates images of the accretion disk for the case when the inclination angle $\iota = 17^{\circ}$ and $80^{\circ}$ and for the different values of BH parameters $\alpha$ and $b$.

One can observe that with an increase of the values of the parameters $\alpha$ and $b$, the photon ring starts to disappear due to the fact that the BH image  with the accretion disk completely or partially overlaps the photon ring. {As $r_{ISCO}$ decreases with increasing parameters $\alpha$ and $b$, the photon ring begins to overlap with the inner edge of the disk.} We hope to compare our results with images of thin accretion disks from future X-ray interferometric missions \cite{Uttley:2019ngm}.

\ \ \
\section{Conclusions\label{Summary}}

{Our results on the optical properties of a rotating black hole in Kaluza-Klein theory described by the parameters of total mass, spin, electric and magnetic charges can be summarized as follows:}

\begin{itemize}
\item The photon motion in the close environment of a rotating black hole in Kaluza-Klein theory is extensively explored. The photon sphere produced by the photons at the last stable orbits shifts towards the central object under the effect of magnetic and electric charges.  

\item For the gravitational lensing in strong field regime, the deflection angle of photons that pass near the boundary of the shadow of the rotating Kaluza-Klein black hole are obtained numerically using the developed ray-tracing code for the photon motion. The obtained results indicate that the deflection angle of photons decrease with the increase of electric and magnetic charges. 

\item Numerous synthetic BH shadows are generated and their properties, together with the light deflection angle, around a Kaluza-Klein black hole are studied. Then synthetic images of Kaluza-Klein black holes are compared with the EHT observations of the M87* SMBH shadow in order to put constraints on the parameters of a Kaluza-Klein BH. The constraint on the upper limit of the dimensionless magnetic (electric) charge of the BH is obtained as $2.52$. The upper limit of the asymmetry parameter of the BH shadow is obtained as $0.05$. 
From the number of BH images generated, the cases $i=17, \alpha=0.7, b=2.00$ and $b=2.34$ are found to most closely match that of the M87* images observed by the EHT collaboration. 

We can conclude that, based on the current precision of the M87* black hole shadow image observation by the EHT collaboration, the shadow observations of Kaluza-Klein BHs are almost indistinguishable from that of Kerr BHs.
Much better observational accuracy than the current capabilities of the EHT collaboration are required in order to place verified constraints on the parameters of modified theories of gravity in the strong field regime. In the future, with improved measurements from the EHT on $\alpha_{*}$, it is likely that the bound on $b_0$ will become more stringent.

\end{itemize}


\section*{acknowledgements}

This work was supported in part
 by Grants F-FA-2021-432, F-FA-2021-510, and MRB-2021-527 of the Uzbekistan Ministry for Innovative Development and by the Abdus Salam International Centre for Theoretical Physics under the Grant No. OEA-NT-01. D.A. was supported by the Teach@T{\"u}bingen and Research@T{\"u}bingen Fellowships.
 T.M. acknowledges also the support from the China Scholarship Council (CSC), Grant No.~2022GXZ005433.

\newpage

\bibliography{gravreferences}

\end{document}